\documentclass[galaxies,article,submit,moreauthors,pdftex,10pt,a4paper]{mdpi}

\firstpage{1}
\articlenumber{x}
\doinum{10.3390/------}
\pubvolume{xx}
\pubyear{2019}
\copyrightyear{2019}
\history{Received: 26 July 2019; Accepted: 18 September 2019; Published: 21 September 2019}

\preto{\abstractkeywords}{\nolinenumbers}

\pdfoutput=1

\Title{Properties of subhalos in the interacting dark matter scenario}

\Author{\'Angeles Molin\'e $^{1,2,}$\orcidA{}, Jascha A. Schewtschenko $^{3}$, Miguel A. S\'anchez-Conde $^{1,2}$, Alejandra~Aguirre-Santaella $^{1,2}$\orcidD{}, Sofía A. Cora $^{4,5,6}$ and Mario G. Abadi $^{7,8}$\orcidF{}}

\AuthorNames{\'Angeles Molin\'e, Jascha A. Schewtschenko, Miguel A. S\'anchez-Conde, Alejandra Aguirre-Santaella, Sofía A. Cora and Mario G. Abadi.}

\address{%
$^{1}$ \quad Instituto de Física Teórica UAM-CSIC, Universidad Autónoma de Madrid, C/ Nicolás Cabrera, 13-15, 28049~Madrid, Spain; miguel.sanchezconde@uam.es (M.A.S.-C.); alejandra.aguirre@uam.es (A.A.-S.)\\
$^{2}$ \quad Departamento de Física Teórica, M-15, Universidad Autónoma de Madrid, E-28049 Madrid, Spain \\
$^{3}$ \quad Institute of Cosmology and Gravitation, University of Portsmouth, Dennis Sciama Building, Portsmouth~PO1 3FX, UK; jascha.schewtschenko@port.ac.uk \\
$^{4}$ \quad Instituto de Astrof\'isica de La Plata (CCT La Plata, CONICET, UNLP),
 Observatorio Astron\'omico, \mbox{Paseo del Bosque}, \mbox{B1900FWA La Plata, Argentina}; sacora@fcaglp.unlp.edu.ar\\
$^{5}$ \quad Facultad de Ciencias Astron\'omicas y Geof\'{\i}sicas,
 Universidad Nacional de La Plata,
 Observatorio Astron\'omico, Paseo del Bosque, B1900FWA La Plata, Argentina\\
$^{6}$ \quad Consejo Nacional de Investigaciones Cient\'ificas y T\'ecnicas (CONICET), Godoy Cruz 2290, \mbox{C1425FQB CABA, Argentina}\\
$^{7}$ \quad Observatorio Astron\'omico, Universidad Nacional de C\'ordoba, Laprida 854, X5000BGR C\'ordoba, Argentina; mario.abadi@unc.edu.ar \\
$^{8}$ \quad Instituto de Astronom\'ia Te\'orica y Experimental (IATE), CONICET-Universidad Nacional de C\'ordoba, C\'ordoba, Argentina}
\corres{Correspondence: angie.moline@uam.es; Tel.: +34-91-299-9873}

\abstract{One possible and natural derivation from the collisionless cold dark matter (CDM) standard cosmological framework is the assumption of the existence of interactions between dark matter (DM) and photons or neutrinos. Such possible interacting dark matter (IDM) model would imply a suppression of small-scale structures due to a large collisional damping effect, even
though the weakly interacting massive particle (WIMP) can still be the DM candidate. Because of this, IDM models can help alleviate alleged tensions between standard CDM predictions
and observations at small mass scales. In this work, we investigate the properties of DM halo substructure or subhalos formed in a high-resolution cosmological N-body simulation specifically run within these alternative models. We also run its CDM counterpart, which allowed us to compare subhalo properties in both cosmologies. We show that, in the lower mass range covered by our simulation runs,
both subhalo concentrations and abundances
are systematically lower in IDM compared to the CDM scenario. Yet, as in CDM, we find that median IDM subhalo concentration values increase towards the innermost regions of their hosts for same mass subhalos. Also similarly to CDM, we find IDM subhalos to be more concentrated than field halos of the same mass.
Our work has a direct application on studies aimed at the indirect detection of DM where subhalos are expected to boost the DM signal of their host halos significantly. From our results, we conclude that the role of halo substructure in DM searches will be less important in interacting scenarios than in CDM, but is nevertheless far from being negligible.}


\begin{document}


\section{Introduction}
\label{sec:intro}

The current standard model of cosmology, $\Lambda$CDM, is based on a
cosmological constant to explain the late-time accelerated expansion of
the Universe and a cold dark matter (CDM) component to account for the
required additional gravitational attraction to form and support the
galaxies and larger structures we observe today~\cite{Aghanim:2018}. In this framework, the
structure of the Universe is formed via a hierarchical, bottom-up
scenario~(see, e.g., \cite{Frenk:2012ph}) with small primordial density perturbations
growing to the point where they collapse into the filaments, walls and
eventually dark matter (DM) halos that form the underlying large-scale-structure filamentary web of the
Universe. The galaxies are embedded in these massive, extended DM halos teeming with self-bound substructure. Any viable
cosmological model has to successfully predict both the abundance and internal properties of these structures and their substructures, and match the observational data on a wide range of scales.
$\Lambda$CDM achieves this challenging feat well on the largest scales~\cite{Netterfield:2001yq,Hinshaw:2003fc,Riess:1998cb,Davis:2007na,Hamuy:2013yta}.
Yet, on small scales tensions have been reported between its predictions
and observations in our local cosmological neighbourhood. The abundance
of DM substructures predicted by numerical simulations of
structure formation exceeds significantly the number of satellite
galaxies observed around the Milky Way and neighbouring Andromeda
galaxy~(see e.g., \cite{Klypin:1999uc, Moore:1999nt}). Various explanation attempts for this and similar discrepancies
such as the ``too big to fail" \cite{Boylan:2011}, ``cusp vs. core" \cite{Dubinski:1991} and ``satellite
alignment" problems~\cite{Goetz:2002vm, Vogelsberger:2012ku} were
brought forward, with some of them attributed to feedback mechanisms
in the baryonic sector that suppressed star formation in such small
halos~(see e.g., \cite{Sawala:2015cdf}), thus leaving them without any observable tracers in the
observational surveys~\cite{Kim:2018}, or alter the DM profiles
within the halos~\cite{Renaud:2013,Rosdahl:2016,Pillepich:2017,Springel:2017,Read:2005,Navarro:1996}. Others turned to alternative models for the DM to account for the lower amount of small subhalos~(see, e.g.,~\cite{Bode:2000gq,Jenkins:2016})) or deviations of their expected properties \cite{Peter:2013,Vogelsberger:2014,Bernal2019}. The latter pathway is not only well motivated, as the properties
of DM has yet remained largely a mystery, but in return also
allows us to use the study of galaxies and their structural properties
as effective probes into the very nature of the elusive nature of the DM particle.

One natural deviation from the collisionless CDM
in the standard model is the assumption of the existence of interactions
between DM and the standard model (SM) particles we know about, in particular, photons or neutrinos~\cite{Boehm:2000gq,Boehm:2001hm,Boehm:2004th}. This
does not only affect, as we show in this article, the formation of DM
structures on small scales, but also provides an explanation for the
exact relic abundance of DM, $\Omega_{\mathrm{cdm}}h^2=0.12011$,
found in the Universe today~\cite{Aghanim:2018}. With such
interactions, DM was in full thermal equilibrium with SM
particles at sufficiently early times and then annihilated into SM
particles until the DM decoupled from the standard sector as the
Universe expanded and cooled down. The cross section needed to retain the observed abundance of DM is surprisingly close to the one expected from the interaction via the weak force in the SM, thus coining the name "weakly interacting massive particles (WIMP)
miracle". Beyond-SM theories provide a variety of
WIMP DM candidates such as the minimal SUSY standard
model with the neutralino and sneutrino and their electroweak scale
interactions~\cite{CHUNG20051}, or the minimal Universal extra
dimension model of the Kaluza-Klein (KK) theory with the first excitation mode of the
gauge field as the lightest KK-particle~\cite{Servant:2003}. When
it comes to the interaction partner, the usefulness of baryons is
limited due to their relatively low abundance in the Universe at any
time and the existing constrains on the cross section with DM from direct
detection experiments. Similarly, charged leptons, whose potential coupling with "leptophilic" DM was initially proposed as an explanation for an excess positron flux from outer space as well as DM direct detection signals~\cite{Fox:2009}, are constrained by e.g. the lack of observations of such interactions in collider experiments~\cite{Fox:2011}. On the other hand, relativistic neutrinos and photons can be found in high abundance in radiation-dominated era of the
early Universe and particle-physics experiments, e.g. particle
colliders, provide only very few constraints on their potential
interaction with DM. Additionally, the cross-section considered in this work is sufficiently low that e.g. the scattering rate of observable cosmic photons on DM halos is negligible as the mean free path of a photon even within the high-density inner regions of large DM halos is still many orders of magnitude larger than these regions themselves.

In our work, we do not pick a specific model, but simply work within an
effective theory, i.e. an effective interaction term between some
unspecified, otherwise sterile DM particles and our SM particles of
choice, photons and neutrinos in the Lagrangian. We will refer to this
model as \emph{interacting dark matter} (IDM). Depending on the actual
type/mass of the mediator in our "black box", this can lead to a
momentum/velocity-dependence of our effective cross-sections but, for
simplicity, we mainly focus in the following on velocity-independent
scenarios. For any given cross section, the DM remains coupled
to the radiation in the early Universe until the latter is diluted
enough as the Universe expands for the DM to become decoupled. As a
result of this coupling, primordial perturbations and, thus, the seeds of
late-time structures, are suppressed within the DM below a
certain scale. This is visible as a cut-off in the linear matter power
spectrum. For a DM–radiation scattering cross section of
$\sigma/\sigma_{\rm Th} = 2
\times 10^{-9} (m_{\rm dm}/{\rm GeV})$ with $\sigma_{\rm Th}$ the Thomson cross section and $m_{\rm dm}$ the DM mass, this characteristic scale is $\sim$100~kpc~\cite{Schewtschenko:2015}
and increases or decreases with the cross section~\cite{Boehm:2001hm,
Sigurdson:2004zp, Mangano:2006mp, Serra:2009uu, Wilkinson:2013kia,
Wilkinson:2014ksa, Cyr-Racine:2013fsa}.

Returning to the premise of using the halo and subhalo population as a
probe into the nature of DM, we can use this
suppression and its consequences for the structure formation to find
bounds on the interaction cross section. Unfortunately, as previously
mentioned, a more direct study of the halo population is difficult
as the distribution of its visible tracers i.e. stars and gas is also
subject to not fully quantified astrophysical processes. Strong lensing
may provide a way to determine the DM profile of larger
halos~\cite{Collett:2017}, but the halos around the
cut-off scale are orders of magnitude smaller. Indirect methods on the
other hand, namely, the detection of the annihilation or decay products
of DM particles, are highly dependent on the statistical and
structural properties of the halo and subhalo population. For instance, the
extragalactic $\gamma$-ray and neutrino signals due DM annihilations,
when estimated via the so-called halo model~\cite{Neyman:1952,Scherrer:1991,Cooray20021}, depend mainly on the DM halo and subhalo
structural properties as well as their abundances~(see e.g.,~\cite{Ullio:2002,Moline:2015,Sanchez-Conde:2014,Ackermann:2015tah,Moline:2017}). Clearly, the considered cosmological model is crucial for such DM searches as different predictions for structure formation on small scales imply different gamma-ray or neutrino signal estimations. Ultimately, this may translate into different constraints on the DM annihilation cross section when compared to those obtained assuming the standard $\Lambda$CDM scenario. In~\cite{Moline:2016}, the isotropic extragalactic signals expected from DM annihilations into $\gamma$-rays and neutrinos were investigated for both IDM and $\Lambda$CDM models using only main halo properties as extracted from DM-only simulations.
In this work, we study the properties of the halo substructure in the same IDM scenario of ref.~\cite{Moline:2016}, for which we now use a set of N-body, DM-only cosmological simulations with higher particle resolution. Due, mainly, to the tidal stripping effects on the subhalo population, describing the subhalo DM density profiles is not a trivial task (see, e.g., the discussion in~\cite{Moline:2017}).  Thanks to our higher resolution simulations, by taking a profile-independent approach, we study IDM halo and subhalo structural properties as a function of the distance to the host halo center and subhalo mass for the first time.  As we explain in this work, one such way to characterize such properties without to assume a given density profile is to consider in the analysis the peak circular velocity $V_{\rm{max}}$  and the radius at which this velocity is attained $R_{\rm{max}}$. In previous works \cite{Schewtschenko:2015,Moline:2016} halo properties were presented as a function of halo mass. 
In order to compare halo and subhalo properties, in this work we also present these properties  as a function of $V_{\rm{max}}$.  On the other hand, in~\cite{Boehm:2014vja} a study about the number of subhalos in a Milky-Way-size halo was performed as a function of $V_{\rm{max}}$. In this work we present such analyses as a function of the distance to the host halo center and subhalo mass.

The work is organized as follows. We briefly summarize the theory behind IDM in section~\ref{sec:idm} followed by a description of our simulations in
section~\ref{sec:simulations}. For both IDM and $\Lambda$CDM models, in
section~\ref{sec:results} we present our results for subhalo
properties such as concentrations, abundances and subhalo radial distributions within the host halos. We finally discuss these results and draw our conclusions in section~\ref{sec:discussion}.

\section{Interacting dark matter}
\label{sec:idm}

In our effective theory of IDM, the interactions between DM and
photons (or alternatively neutrinos) result in additional terms in the
 equations  governing the evolution of the cosmic components (see, e.g.,~\cite{Ma:1995}),
\begin{eqnarray}
 \dot \theta_b &=& k^2 \psi - \mathcal{H} \theta_b + c_s^2 k^2 \delta_b
- R^{-1} \dot \kappa \left(\theta_b - \theta_\gamma \right) ~,\\
 \dot \theta_\gamma &=& k^2 \psi + \left( \frac 1 4 \delta_\gamma -
\sigma_\gamma \right) k^2 \delta_b - \dot \kappa \left(\theta_\gamma -
\theta_b \right) - C_{\gamma-\mathrm{DM}} ~,\\
 \dot \theta_{\mathrm{DM}} &=& k^2 \psi - \mathcal{H}
\theta_{\mathrm{DM}} - C_{\mathrm{DM}-\gamma} ~,
\end{eqnarray}
where $\psi$ is the gravitational potential, $\mathcal H = a H$ is the
conformal Hubble rate, $c_s$ is the baryon sound speed and $\delta$, $\theta$ and $\sigma$ are the density, velocity divergence and anisotropic stress potential respectively, associated with the
 baryon (b), photon ($\gamma$) and DM fluid. For the electromagnetic interactions (EM) in the
SM, the first two equations include terms with the Thomson scattering rate $\dot \kappa \equiv a
\sigma_{\mathrm{Th}} c n_e$, where $c$ the speed of light and $n_e$ the density of free electrons (the scale factor $a$, appears since the derivative is taken with respect to conformal time). The ratio of the baryon to photon density,  $R \equiv (3/4)(\rho_b / \rho_\gamma)$, is
a pre-factor to ensure momentum conservation. $C_{\mathrm{DM}-\gamma}$
and $C_{\gamma-\mathrm{DM}}= -S^{-1} C_{\mathrm{DM}-\gamma}$ are the new
interactions terms that have to be added to include interactions between
DM and the cosmic photon background with $S \equiv (3/4)(\rho_{\mathrm{DM}}
/ \rho_\gamma)$ as the scaling of the counter term in the momentum and $\rho_{DM}$ is the dark matter energy density. Analogous to the EM
interaction,
\begin{equation}
C_{\mathrm{DM}-\gamma}= \dot \mu \left(\theta_{\mathrm{DM}} -
\theta_{\gamma} \right)
\end{equation}
depends on the new interaction rate $\dot \mu \equiv a
\sigma_{\mathrm{DM}-\gamma} c n_{\mathrm{DM}}$. Here
$\sigma_{\mathrm{DM}-\gamma}$ is the elastic scattering cross-section
between DM and photons while $n_{DM} = \rho_{DM}/m_{DM}$ is the
DM number density. For the DM-neutrino interactions, similar modifications can be
added. In \cite{Wilkinson:2013kia} an implementation of
these modified Euler equation for the \texttt{CLASS} Boltzmann solver was presented.
We are using this work to calculate the linear evolution of the Universe up
to the point (in this work at redshift $z=127$) where we switch to simulations to also cover the full
non-linear evolution and resulting structure formation accurately (for more details see also~\cite{Boehm:2014vja}).


\section{Simulations}
\label{sec:simulations}
%
%

For this work, we calculate  the  non-linear  evolution  of  the
matter  distribution using a suite of cosmological DM-only simulations.
This includes both simulations of single-resolution periodic volumes of
$100$~Mpc  as well as zoom-in simulations which focus on representative
sub-volumes to improve the maximum resolution for a subset of the
obtained DM structure samples.\\
We perform these simulations with  the parallel Tree-Particle Mesh
$N$-body code, \texttt{P-Gadget3}~\cite{Springel:2005mi} for both a standard,
collision-less CDM and a $\gamma$CDM model with a cross section $\sigma/\sigma_{\rm Th} = 2
\times 10^{-9} (m_{\rm DM}/{\rm GeV})$.  This value is (roughly) the upper bound obtained in previous works from satellite number counts of Milky-Way-size halos~\cite{Boehm:2014vja,Schewtschenko:2015}.  In~\cite{Escudero:2018} a more conservative constraint is claimed using measurements of the ionization history of the Universe at several redshifts, results from N-body simulations and recent estimates of the number of Milky Way satellite galaxies. However, the approach implemented  can generate large uncertainties since the presence of low-mass
subhalos in galactic halos which simulations can not resolve and extrapolations are necessary to obtain the results. Note that whereas larger cross sections would erase most of the observed substructure, smaller cross sections would imply results in between CDM and IDM.
The simulations begin at a redshift of $z = 127$ (the DM--radiation
interaction rate is negligible at all times afterwards). For the
initial conditions we use the same cosmology (WMAP7), random phases and
second-order Lagrangian perturbation theory (2LPT) method~\citep{Jenkins:2010} as the \texttt{APOSTLE}
project \citep{Fattahi:2016MNRAS} and our previous studies of the impact
of IDM on galactic
substructures~\cite{Boehm:2014vja}. After having performed the
full-volume run for both standard CDM and $\gamma$CDM with a particle
mass $m_{\rm Part} = 1.96 \times 10^{8}~{\rm M}_\odot/h$ and a comoving
softening length $l_{\rm soft} = 2.7~\textrm{kpc}$, we identify the DM
structures within using the \texttt{Rockstar} halo
finder~\cite{Behroozi:2012}. All halo properties are determined for spherically overdense regions with a density of 200 times the critical density of the Universe at present, $\rho_{c}$.
With these results, a cubic sub-volume is chosen at $z=0$ with a side length of $14$ Mpc/h that reproduces the overall halo mass function on the mass scales covered by it. A $1~\rm{Mpc}$ wide margin is added and the resulting volume traced back
to the initial redshift. We checked that the sub-volume thus constructed
is still convex in these Lagrangian coordinates. This ensures that the
progenitors of the structures within the targeted region evolve well
within the high-resolution region, when the resulting volume is re-run using a
zooming technique~\cite{Onorbe:2014} with $m_{\rm Part} = 4.85 \times 10^{5}~{\rm
M}_\odot/h$ and $l_{\rm soft} = 860~\textrm{pc}$ in the targeted region.

\begin{table}[H]
\caption{Most relevant parameters of Box and LGs simulations, together with their corresponding halo and subhalo abundances. Columns 2-4 indicate the box size $L_{\rm sim}$, the particle mass $m_{\rm Part}$, and the comoving softening length $l_{\rm soft}$. Rest of columns provide the total number of subhalos $N_{\rm sub, IDM/CDM}$, and halos $N_{\rm h, IDM/CDM}$ for each cosmological model. We remind that there are 4 LGs in each case.}
\centering
\begin{tabular}{cccccccc}
\toprule
\textbf{}   & \textbf{$L_{\rm sim}$}   & \textbf{$m_{\rm Part}$}  & \textbf{$l_{\rm soft}$} & \textbf{$N_{\rm sub, IDM}$} & \textbf{$N_{\rm sub, CDM}$} & \textbf{$N_{\rm h, IDM}$} & \textbf{$N_{\rm h, CDM}$}\\
\midrule
Box		& $100$~Mpc 			& $1.96 \times 10^{8}~{\rm M}_\odot/h$  & $2.7~\textrm{kpc}$ & 17481  & 27973 & 125704 & 197208\\
LGs		& $15$ Mpc/h   		& $4.85 \times 10^{5}~{\rm M}_\odot/h$  & $860~\textrm{pc}$  & 1606  & 11092 & 10513 & 40874\\
\bottomrule
\end{tabular}
\label{tab:1}
\end{table}

Throughout this work, we use the term \textit{Box} to refer to the full-volume simulation ($100$~Mpc) at $z=0$ for each cosmology. The zoom re-simulations model four Local Groups (\textit{LGs} hereinafter). We filter the results to pick only those halos that are well within the higher resolution region, namely inside a $\sim 2.1$ Mpc/h radius at $z=0$. This is done in order to avoid boundary affects, such has halos that consist partly of higher-mass particles, which are ignored here. The total number of halos and subhalos found in both Box and LGs simulations is given in Table~\ref{tab:1}, together with the most relevant parameters of these simulations.

\section{Results}
\label{sec:results}

As mentioned, IDM exhibits a linear matter spectrum different to the one of CDM~\cite{Boehm:2001hm, Sigurdson:2004zp,Mangano:2006mp,Serra:2009uu,Wilkinson:2013kia,Wilkinson:2014ksa,Cyr-Racine:2013fsa}. The IDM matter power spectrum features a cut-off around a smooth scale of $\sim 100$ kpc for the cross section that we are considering in this work ( $\sigma/\sigma_{\rm Th} = 2\times 10^{-9} (m_{\rm DM}/{\rm GeV})$).
Therefore a suppression of the number of halos below the scale of those
hosting dwarf galaxies is expected (i.e. for halo masses below $ \sim 10^{10}~{\rm M}_\odot/h$). In addition,  such linear matter power spectrum impacts the structural halo properties, such as shape, spin, density profile and halo concentrations~\cite{Schewtschenko:2015,Boehm:2014vja,Moline:2016}. 
In this section, we show the results we found for halo and subhalo concentrations in our simulations, as well as subhalo abundances.

\subsection{Halo concentrations}
\label{sec:halo concentrations}
We consider two different definitions for the concentration parameter. The first and more standard is $c_{\Delta} \equiv R_{\rm{vir}}/r_{-2}$, i.e. the ratio between the halo virial radius, $R_{\rm{vir}}$, and the radius $r_{-2}$ at which the logarithmic slope of the DM density profile $\frac{d \log\rho}{d \log\,r}=-2$. The other definition has the advantage of being  independent of the adopted DM density profile and the particular definition used for the virial radius \citep{Diemand:2007qr,Diemand:2008in,Springel:2008cc}:
\begin{equation}
c_{\rm V}=\frac{\bar{\rho}(R_{\rm{max}})}{\rho_{c}} = 2\left( \frac{V_{\rm{max}}}{H_{0} \, R_{\rm{max}}}\right)^{2}\,,
\end{equation}
where $\bar{\rho}$ is the mean physical density within $R_{\rm{max}}$ and $H_{0}$ the Hubble constant.  At a given $V_{\rm{max}}$, the concentration provides an alternative
measure of the characteristic density of a halo.

Assuming an NFW profile \citep{Navarro:1995iw, Navarro:1996gj}, the relation between $c_{\rm V}$ and $c_\Delta$ is given by \citep{Diemand:2007qr}
\begin{eqnarray}
c_{\rm V}&=&\left( \frac{c_{\Delta}}{2.163}\right)^{3}\,\frac{f(R_{\rm{max}}/r_{s})}{f(c_\Delta)}\,\Delta ~,
\label{eq:cvcD}
\end{eqnarray}
where $f(x)= \mbox{ln}(1+x)-x/(1+x)$ and $r_{s}=r_{-2}$ is the scale radius. For spherical (untruncated) halos\footnote{Which are not affected by tidal forces.} with a virial mass $M_{\Delta}$ and virial radius $R_\Delta$ at redshift $z=0$, we have
\begin{equation}
\label{eq:D}
M_{\Delta}  = \frac{4\pi}{3} \, R_{\Delta}^{3} \, \rho_c \, \Delta ~.
\end{equation}
where $\rho_c$ is the critical density of the Universe at present, $\Delta$ is the overdensity factor that defines the halos and
$r_{\Delta}$ is its virial radius.

 Using our set of simulations, both Box and LGs for IDM and CDM models, we obtain the medians of $c_{\rm V}$ and $c_{\Delta}$. The latter was found by applying the $c_{\rm V}$-$c_\Delta$ relation of Eq.~(\ref{eq:cvcD}) to the $c_{\rm V}(V_{\rm max})$ values found for every halo in the simulations. We adopt $\Delta = 200$ as the value for the overdensity to define the halos. For Box, we applied a restriction on halo maximum circular velocity such that only halos with $V_{\rm{max}} > 60$~km/s are included; in the case of the LG data set this restriction is set at $V_{\rm{max}} > 10$~km/s. Both criteria are adopted in order to avoid resolution issues in the determination of $c_{\rm V}$ at the smallest scales resolved by the simulations. We have grouped halos in bins of $V_{\rm{max}}$ and have obtained the medians of $c_{\rm V}$. For both the LGs and Box simulations, similar bin sizes were chosen to cover the entire $V_{\rm{max}}$ range, $\sim 10$~km/s $< V_{\rm{max}} <  10^3$~km/s. For each cosmology, we consider 5 bins in LGs and 9 bins for Box simulations.

\begin{figure}
\includegraphics[width=0.5\textwidth]{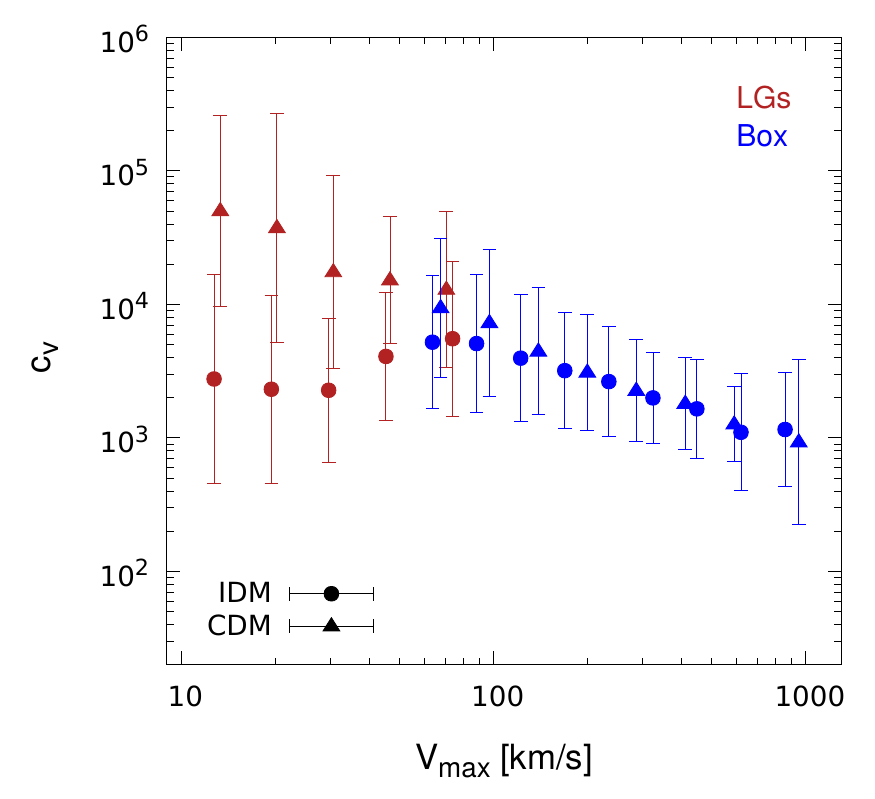}
\includegraphics[width=0.5\textwidth]{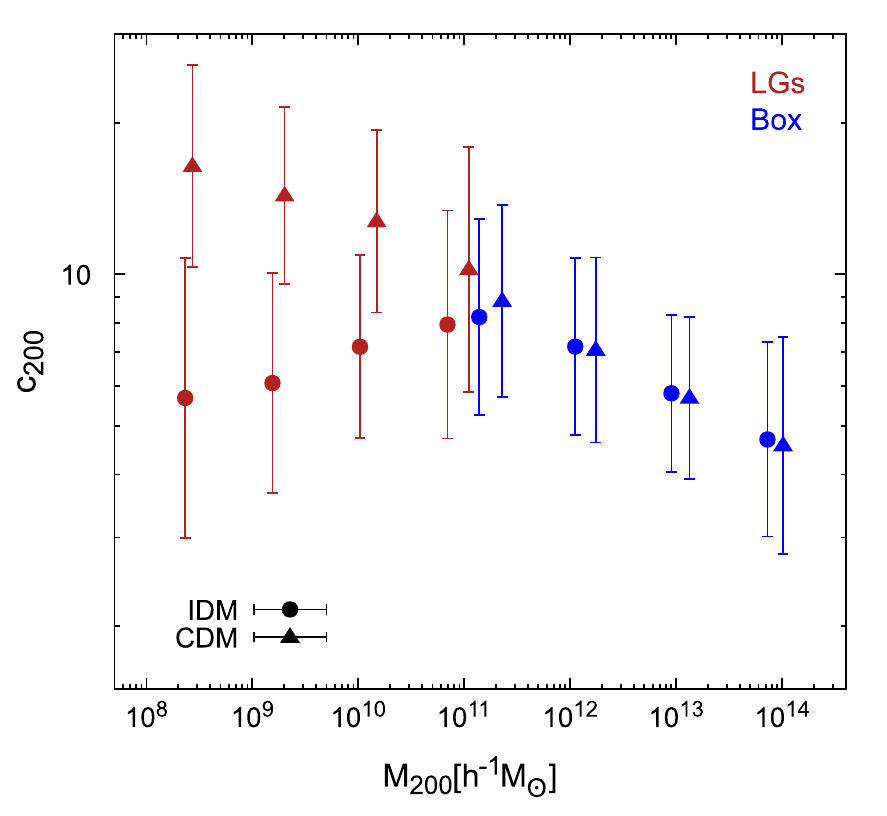}
\caption{Median halo concentrations and $1\sigma$ errors as found in our set of simulations, Box (blue) and LGs (red), at $z=0$. (\textbf{a}) Left panel: median $c_{\rm V}$ values as a function of $V_{\rm{max}}$.  (\textbf{b}) Right panel: $c_{200}$ as a function of $M_{200}$. In both panels, the circle symbols refer to the IDM simulations whereas the triangles to CDM.}
\label{fig:halo-conc}
\end{figure}

 In Fig.~\ref{fig:halo-conc} we show halo concentration values and corresponding $1\sigma$ standard deviation (left panel) as found in Box (blue) and the four LGs (red) simulation runs. Left and right panels show, respectively, results for both median $c_{\rm V}(V_{\rm{max}})$ and $c_{200}(M_{200})$ values, the latter in bins of the halo mass $M_{200}$ (4 for both Box and LGs data), calculated using Eq.~(\ref{eq:D}) and covering a mass range of $\sim 10^{8}~{\rm M}_\odot/h < M_{200} < 10^{14}~{\rm M}_\odot/h$. In order to have truly isolated  halos, the so-called "field halos", in our analysis, we only considered halos that do not have another massive neighbour (defined as more than half the mass of the halo under consideration) located within a distance of 1.5 times its virial radius, $R_{200}$. In order to compare IDM and CDM subhalo concentrations one-to-one, we also include in Fig.~\ref{fig:halo-conc} the corresponding CDM concentrations.
First, it is worth noting that Fig.~\ref{fig:halo-conc} shows an excellent agreement between the concentrations values found in both Box and LGs at the scale where the simulations overlap. Also, as expected, both IDM and CDM yield similar results at large halo masses, while we derive significantly lower median concentration values below halo masses $\sim 10^{11}~{\rm M}_\odot/h$ in the case of IDM compared to CDM. Interestingly, this decrease of concentration values is similar to that found in WDM simulations, an effect that has been explained as being due to the delayed formation time of low-mass halos~\cite{Lovell:2012}. In addition, similar analysis for $c_{200}$ was performed in~\cite{Schewtschenko:2015} and \cite{Moline:2016} where also the dependence with redshift was presented. Our results are in good agreement with such previous ones at $z=0$. As we explained above, such results for the concentration-mass relation, $c_{200}(M_{200})$, were obtained from $c_{V}(V_{\rm max})$ (see Eq.~\ref{eq:cvcD}). In this way we double check previous results for IDM halo concentrations where a NFW profile was assumed. At late times, interacting DM models become (effectively) non-collisional for the cross section studied here, in the same way that the free-streaming in WDM models becomes negligible at low redshifts. Therefore, the observed lower IDM concentration values at small halo masses also originate from the later collapse of DM halos in these models.

\subsection{Subhalo concentrations}
\label{sec:sub-c}

The same analysis in $V_{\rm max}$ and subhalo mass, $m_{200}$, bins was performed for $c_{\rm V}$ and $c_{200}$  subhalo concentrations respectively.  In this case for Box, 8 bins are considered to cover the $V_{\rm max}$ range and 5 for $m_{200}$. We applied a restriction on subhalo maximum circular velocity such that only subhalos with $V_{\rm{max}} > 60$~km/s are included; for the \textit{LGs},  this restriction is set at $V_{\rm{max}} > 10$~km/s considering just 3 bins for both $V_{\rm max}$ and  $m_{200}$  in order to obtain the median concentration values with a good subhalo statistics. From the results of Box and LGs simulations together, the $V_{\rm max}$ range covered  is $10 < V_{\rm{max}} <  500$~km/s   in each cosmology.

\begin{figure}
\includegraphics[width=0.50\textwidth]{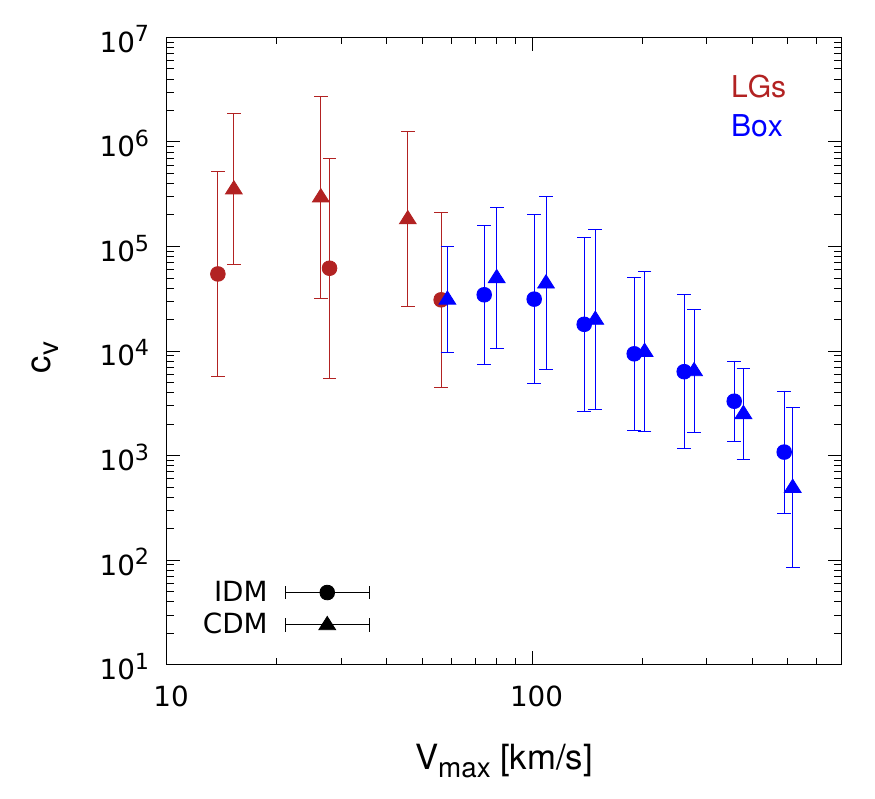}
\includegraphics[width=0.50\textwidth]{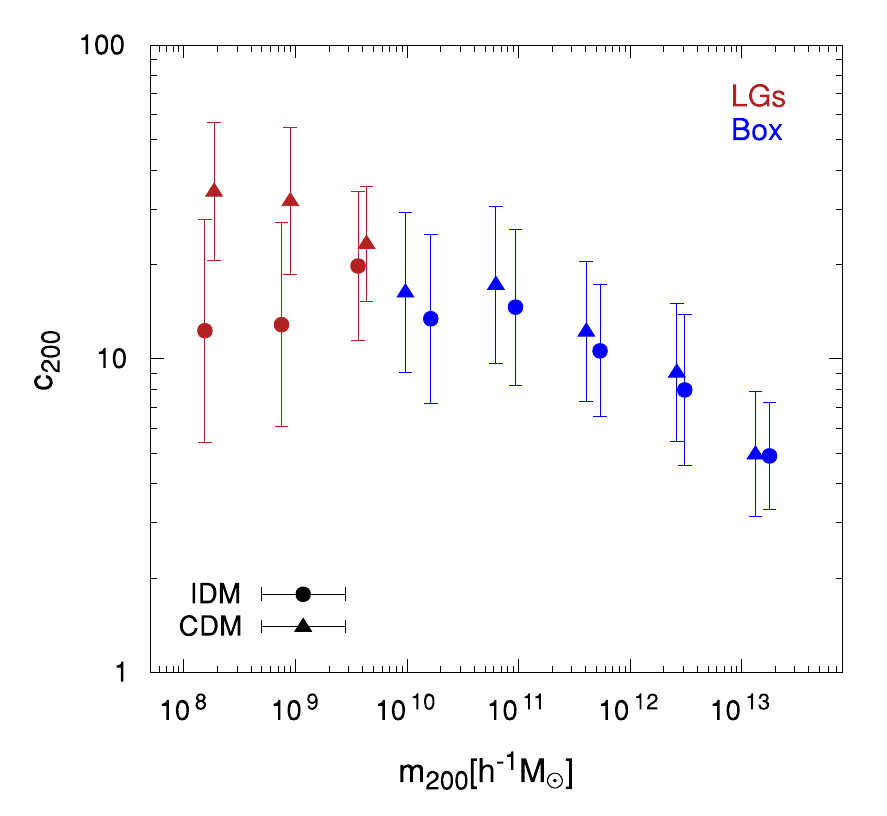}
\caption{Median subhalo concentrations and $1\sigma$ errors as found in our set of simulations, Box (blue) and LGs (red), at $z=0$. The circle symbols represent the results from the IDM simulations, whereas the triangle symbols correspond to the CDM results. (\textbf{a}) Left panel: the median $c_{\rm V}$ as a function of $V_{\rm{max}}$. (\textbf{b}) Right panel: $c_{200}$ as a function of $m_{200}$ as obtained using Eqs.~\ref{eq:cvcD} and~\ref{eq:D} for every subhalo in the simulations.
}
\label{fig:sub-conc}
\end{figure}

In the left panel of Fig.~\ref{fig:sub-conc}, we depict median $c_{\rm V}(V_{\rm{max}})$ values and corresponding $1\sigma$ errors as found in  Box (blue) and the four LGs (red). The right panel shows the results for $c_{200}(m_{200})$. As in Fig.~\ref{fig:halo-conc}, we also include the corresponding CDM concentrations. As it can be seen, the medians of $c_{\rm V}$ ($c_{200}$) in both cosmologies are similar for $V_{\rm{max}} > 60$~km/s ($m_{200}> 2 \times 10^{9}$~M$_{\odot}$/h), while there is a significant departure between them at lower $V_{\rm{max}}$ ($m_{200}$) values. Unfortunately, the simulations have a limited mass resolution and subhalo statistics in that range, which translates into large $1\sigma$ errors and, as a consequence, our results are not conclusive. Yet, they provide a consistent picture of the subhalos' concentration behaviour at small $V_{\rm{max}}$ ($m_{200}$) values, IDM subhalos exhibiting lower concentrations than CDM subhalos in the mentioned $V_{\rm{max}} < 60$~km/s range.

\begin{figure}
\centering
\includegraphics[width=0.49\textwidth]{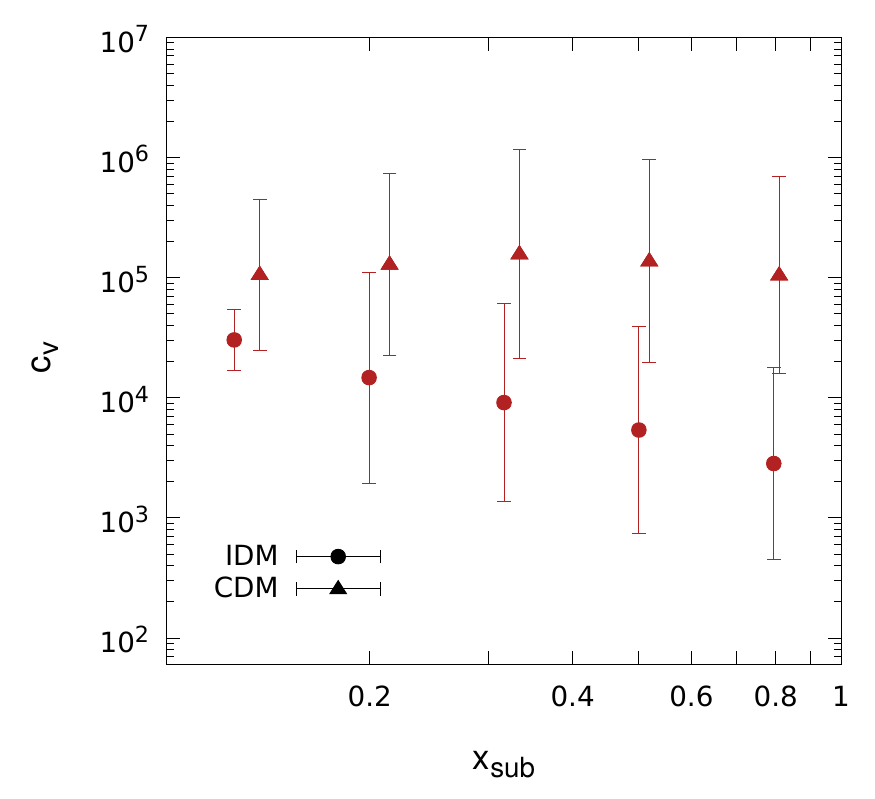}
\includegraphics[width=0.49\textwidth]{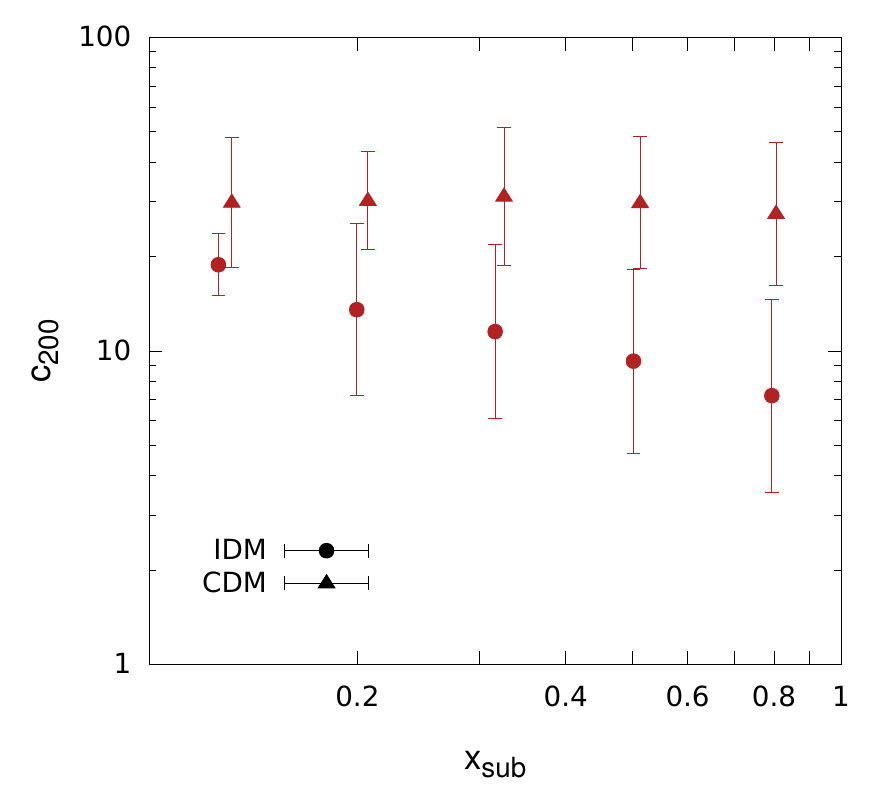}
\caption{Median subhalo concentrations and $1\sigma$ errors as a function of $x_{\rm sub}$, i.e., the distance to the center of the host halo normalized to $R_{200}$. We show results for $c_{\rm V}$ (left) and $c_{200}$ (right) as derived from our set of LGs simulations.}
\label{fig:x-c}
\end{figure}

Assuming a CDM framework, previous works have shown that the subhalo concentration depends not only on the mass of the subhalo but also on the distance to the center of its host halo~\cite{Diemand:2008in,Pieri:2009je,Moline:2017}.
In order to know  if the same behaviour is found for IDM subhalos,
Fig.~\ref{fig:x-c} depicts, for the LGs, the medians and $1\sigma$ errors of $c_{\rm V}$ (left panel) and $c_{200}$ (right panel) as a function of the distance from the host halo center in units of $R_{\rm 200}$.  As before, we also include in the figure our results for the CDM case which  are in good agreement with previous ones presented in \cite{Moline:2017}. Median IDM subhalo concentration increases towards the center of the host halo more significantly than in the CDM case.  Yet, for each considered radial bin, IDM concentrations are significantly and {\it consistently} lower than CDM ones. Such effects could be understood by studying in detail the properties of both CDM and IDM subhalos at infall. This study is beyond the scope of this paper and will be presented elsewhere. Again, large error bars prevent us from extracting firm conclusions and, thus, we will not propose any parametric fits to the data in this paper.  However, this is an interesting qualitative result that points to a significantly different distribution of subhalo concentrations inside the host halo in the IDM scenario compared to CDM.

\begin{figure}
\includegraphics[width=0.50\textwidth]{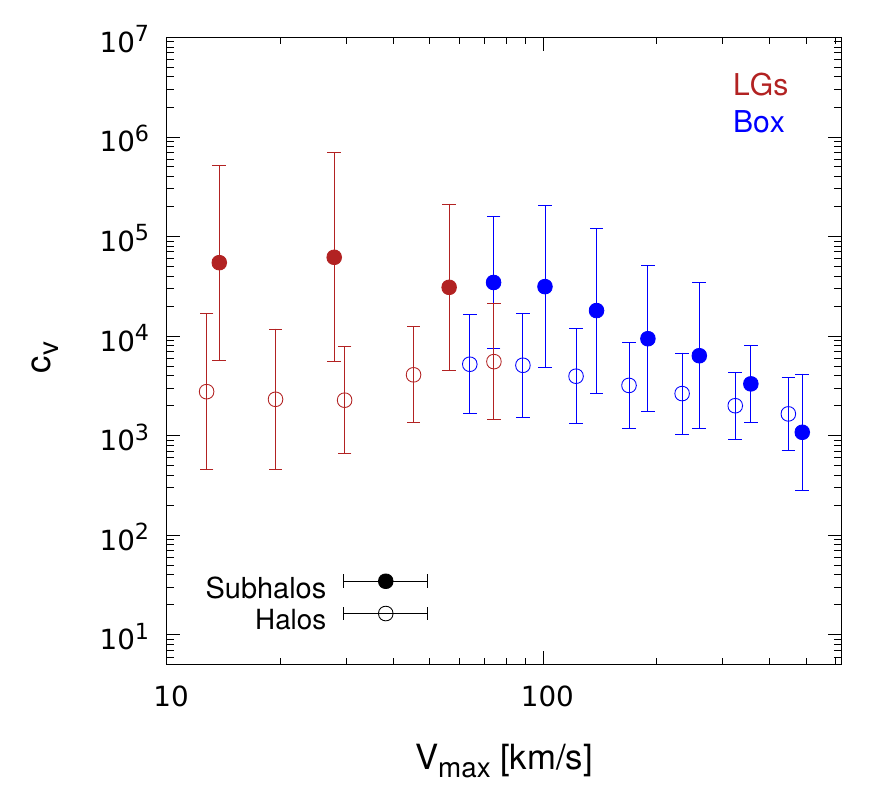}
\includegraphics[width=0.50\textwidth]{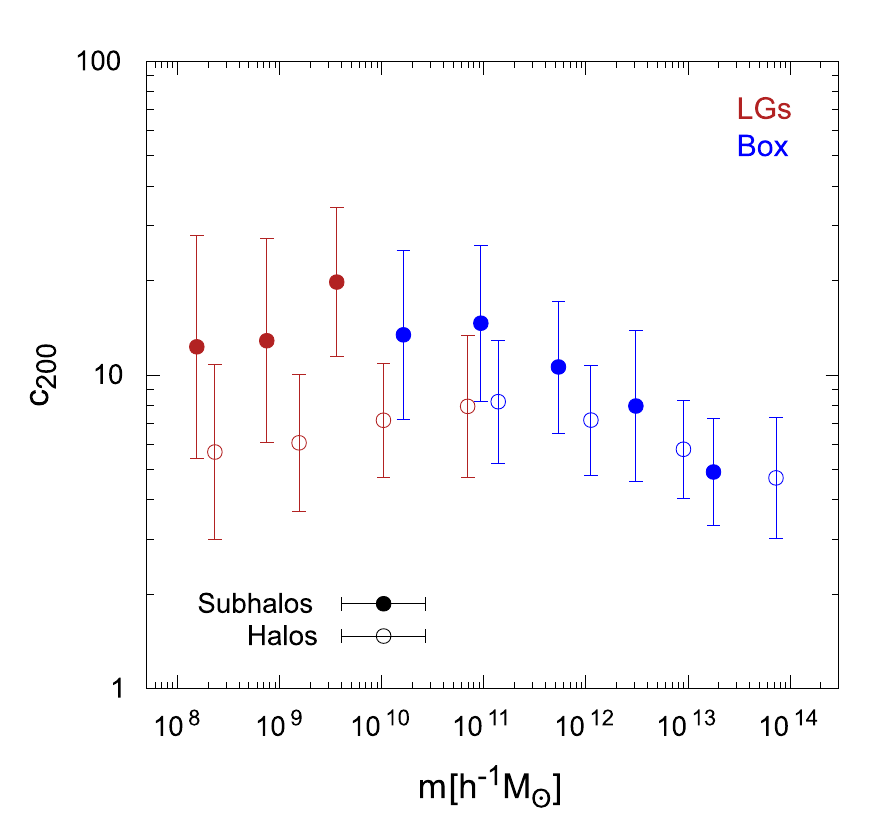}
\caption{Left panel: Median halo (open circles) and subhalo (filled circles) $c_{\rm V}$ concentration values and corresponding $1\sigma$ errors, as a function of $V_{\rm max}$, as found in our set of simulations for IDM at $z=0$: Box (blue) and LGs (red). Right panel: the same but for $c_{200}$ as a function of $m_{200}$.}
\label{fig:halo-subh-c}
\end{figure}
In the standard CDM cosmological framework, it is well established from simulations that subhalos are more concentrated than field halos of the same mass~\cite{Ghigna:1999sn,Bullock:1999he,Moore:1999nt,Ullio:2002, Diemand:2008in,Diemand:2009bm,Pieri:2009je,Bartels:2015uba,Moline:2017}. It might not be the case in the IDM model, indeed the mean subhalo concentration values (see Fig.~\ref{fig:sub-conc}) fall within the  values of halos concentrations studied in previous works for CDM. However, from Fig.~\ref{fig:halo-conc} we see that the IDM halos exhibit lower concentrations compared with the halo concentrations in CDM of the same mass and then differences are expected between the concentrations of subhalos and their hosts in the interacting models.  In Fig.~\ref{fig:halo-subh-c} we shape such differences between halos and subhalos in the IDM scenario by comparing their median $c_{\rm V}$ ($c_{200}$) values and $1\sigma$ errors as a function of $V_{\rm max}$ ($m_{200}$) as found in our set of simulations. Analogously to what occurs in CDM, we obtained that also in IDM models subhalos with mass $m_{200} < 10^{11}$~M$_{\odot}$/h tend to be more concentrated than their host halos. As in previous cases above, a more quantitative statement about the observed trend is nevertheless not possible for the moment, given the relatively large uncertainties involved in our study.

\subsection{Subhalo abundances}
\label{sec:SMF}

As mentioned, DM interactions lead to a matter power spectrum different from the one in CDM. This matter power spectrum features a cut-off around a smooth scale of $\sim 100$ kpc, and therefore a suppression of the number of halos in the lower mass range. The impact of such IDM initial matter power spectrum on the abundance of halos was studied in~\cite{Schewtschenko:2015,Moline:2016}, where a
comparison with the standard CDM result was also presented. A suppression of the number of low-mass halos with masses below $M_{200} \sim 10^{11} \, h^{-1} \, M_\odot$ was found, which became particularly significant at the smallest considered halo masses. In this section, we will complement these previous studies by using our set of IDM simulations to obtain the first results for subhalo abundances. We will do so in a broad subhalo mass range, i.e., $[2 \times 10^{6},10^{12}]~{\rm M}_\odot/h$.

\begin{figure}
\centering
\includegraphics[width=0.9\textwidth]{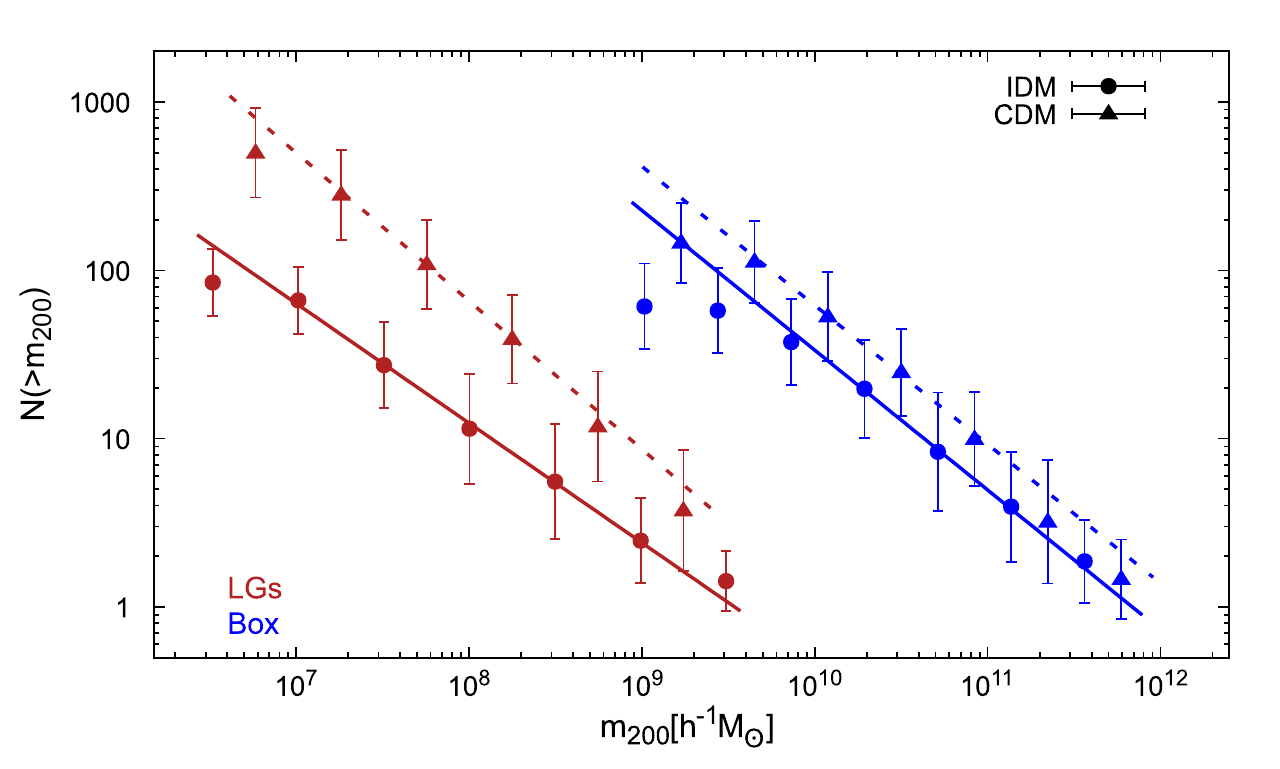}
\caption{Cumulative number of subhalos, $N(\rm > m_{200})$, as a function of subhalo mass, $m_{200}$, in the case of IDM (circle symbols) and CDM (triangles) as obtained from Box (blue) and LGs (red) simulations at $z=0$. We also show the corresponding fits using Eq.~\ref{eq:csmf} with the best-fit parameters reported in Tab.~\ref{tab:2} (solid colored lines).}
\label{fig:CSMF}
\end{figure}

In Fig.~\ref{fig:CSMF}, we show the cumulative number of subhalos, $N(\rm > m_{200})$, as a function of subhalo mass, $m_{200}$, for both IDM and CDM scenarios and for both Box and LGs.  Then, we consider all subhalos residing in halos with $M_{\rm h} > 3 \times 10^{13}~M_{\odot}/h$ for Box, and $3 \times 10^{11}~M_{\odot}/h < M_{\rm h} < 1.4 \times 10^{12}~M_{\odot}/h$ for LGs. These ranges allow us to have more than 30 subhalos per host in both cosmologies and both simulation sets. For each halo, we calculate the cumulative number of subhalos by adopting 100 subhalo mass bins and by finding the mean for each subhalo mass bin over all the main halos in the corresponding simulation. In the same Fig.~\ref{fig:CSMF}, we also show in solid lines the result of fitting the data with the following parametric expression:
\begin{eqnarray}
 N (> m_{200}) = \beta\, m^{\gamma}_{200}
\label{eq:csmf}
\end{eqnarray}
This fitting function follows previous works that calculated the cumulative subhalo mass function from N-body cosmological simulations, and where the subhalo mass function was found to obey a power law $dN/dm \, \propto  m ^{-\alpha}_{200}$.~\cite{Diemand:2007qr}. Both the normalization factor, $\beta$, and the slopes $\gamma=-\alpha+1$, will depend on the adopted cosmological model. In Tab.~\ref{tab:2}, we report the best-fit values we found in our simulations for $\gamma$ and $\beta$, both for CDM and IDM scenarios. Using the LG data, the fits works well for the subhalo mass range $[0.59, 3.39] \times 10^{10}$ M$_\odot/h$ and $[1.19, 9.66] \times 10^{10}$ M$_\odot/h$ for Box, in both cosmological models.
\begin{table}[H]
\caption{Best-fit parameters and $\chi^{2}$  values for the cumulative subhalo mass function given in Eq.~\ref{eq:csmf} according to our data. We show results for both IDM and CDM as obtained from our LGs and Box simulations. 
}
\centering
\begin{tabular}{ccccccc}
\toprule
\textbf{}	& \textbf{$\gamma^{LGs}$}  	 &\textbf{$\beta^{LGs}$}  &\textbf{$\chi^{2,LGs}$}   & \textbf{$\gamma^{Box}$}  & \textbf{$\beta^{Box}$} &  \textbf{$\chi^{2,Box}$}\\
\midrule
 IDM & $-~0.71$  & 6.04 $\times 10^{6}$  & 0.27 & $-~0.83$ &  6.74 $\times 10^{9}$ & 0.068\\
 CDM & $-~0.88$  & 7.22 $\times 10^{8}$  & $0.19$ & $-~0.83$   & 1.10 $\times 10^{10}$  & 0.36\\
\bottomrule
\end{tabular}
\label{tab:2}
\end{table}

As it can be seen in Fig.~\ref{fig:CSMF} and in Tab.~\ref{tab:2}, in the case of the LGs the normalization of the cumulative subhalo mass function in the IDM case is significantly lower than that of CDM subhalos.
More precisely, we find that mean $N(\rm > m_{200})$ values for IDM subhalos are almost a factor $\sim 10$ lower than those of CDM for subhalos in the range $10^{7}~M_{\odot}/h < m_{200} <  10^{8}~M_{\odot}/h$, this factor decreasing towards large subhalo masses. In Box, which covers comparatively larger halo masses, the differences among the two considered cosmologies are not statistically significant anymore. Indeed, all these results are as expected. As discussed above, the particular differences between the IDM and CDM initial matter power spectra lead to a suppression of smaller structures in the former case with respect to the latter, an effect that must become more evident in the LGs compared to Box, as the former simulations resolve smaller subhalo masses.
\begin{figure}
\centering
\includegraphics[width=0.9\textwidth]{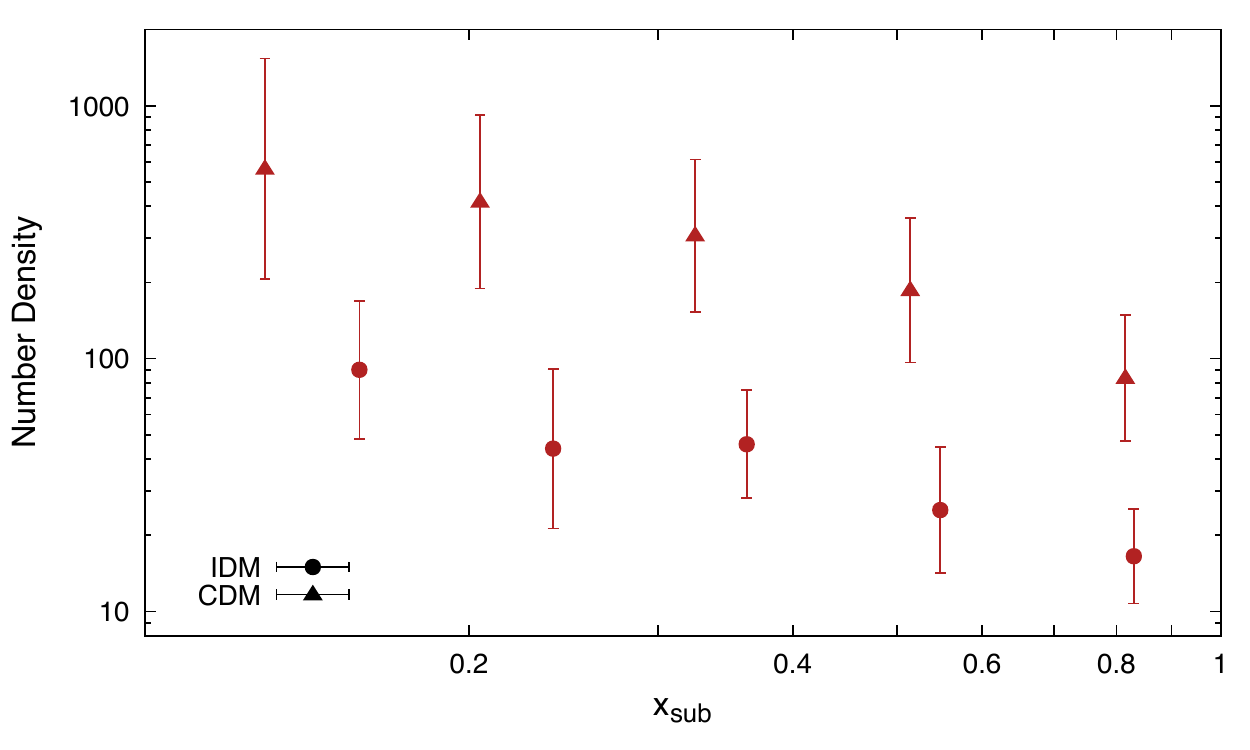}
\caption{Number density of subhalos as a function of distance to the host halo center, $x_{\rm sub}=r_{\rm sub}/R_{\rm 200}$. We show results for both IDM (circle symbols) and CDM (triangles). Both cases refer to the LGs simulation set; see text for details.
}
\label{fig:SRD}
\end{figure}
Finally, we also studied the radial dependence of the number of subhalos in the IDM case, and compared it to the more standard CDM subhalo radial distribution. We did so only for the LGs, since high resolution simulations are necessary to perform this kind of analysis. Indeed, we checked that the statistics in the Box simulation is not sufficient to properly perform the work.  Fig.~\ref{fig:SRD} depicts mean values and corresponding $1\sigma$ errors of the number density as a function of the distance from the center of the host halo (in units of its $R_{\rm 200}$) for halos with  [0.5~--~1] $\times 10^{12}~M_{\odot}/h$. As it can be seen, the radial number density of IDM subhalos increases towards the center of the host halo as in the CDM case but is significantly lower than the latter at all host radii.

\section{Summary and discussion}
\label{sec:discussion}

We have investigated DM subhalo properties in models where the linear matter power spectrum is suppressed at small scales due to DM interactions with radiation (photons or neutrinos).  We do so by making use of N-body cosmological simulations, which are known to be a crucial tool to study the properties of DM structures. More precisely, we use data from our own set of simulations, described in Sec.~\ref{sec:simulations}. The runs are performed in both the standard CDM paradigm and in the IDM scenario, where the latter assumes interactions of DM with photons.\footnote{We do not include the case of DM-neutrino interactions, yet the results are expected to be similar to those presented in this work; see discussions e.g. in~\cite{Boehm:2014vja,Schewtschenko:2015}.} This allows us to compare DM halo and subhalo properties as found in both cosmologies. Since the main impact of the DM-photon interactions on structure formation occurs mainly at small scales, we use data not only from a large simulation box (100 Mpc) but also high-resolution zoom-in simulations of four Local Groups.

First, in sections~\ref{sec:halo concentrations} and \ref{sec:sub-c}, we studied, respectively, halo and subhalo concentrations as a function of halo/subhalo mass (and, alternatively, $V_{\rm max}$).
Both for halos and subhalos we observed a significant reduction of the concentrations in the lower mass range (or, alternatively, small $V_{\rm max}$ values). Our result for halos confirm the findings of previous works, e.g. \cite{Schewtschenko:2015,Moline:2016}, while this is the first time that the concentration of IDM subhalos was studied. This decrease of concentration values is expected and originates from the later collapse of low-mass DM halos and subhalos in  IDM cosmologies, similarly to that observed in WDM simulations~\cite{Lovell:2012}.

Also in section~\ref{sec:sub-c}, we studied subhalo concentrations as a function of the subhalo distance to the host halo center. As in the CDM framework, we found that the median subhalo concentration values increase towards the innermost regions of the host for subhalos of the same mass. Yet, we obtained significantly lower median concentrations in the IDM case with respect to CDM at all radii (see Fig.~\ref{fig:x-c}). Limitations in the number of subhalos prevent us from quantifying this effect more in detail, thus it seems robust in clearly present in our data. New N-body cosmological simulations with improved resolution will be needed in order to perform a more exhaustive analysis in this direction.

In addition, when comparing our results for IDM halos and subhalos of the same mass, we conclude that  in these IDM models the subhalos are more concentrated than field halos (see Fig.~\ref{fig:halo-subh-c}), similarly to what found for CDM, e.g.~\cite{Moline:2017}.

Finally, we also presented in section~\ref{sec:SMF} our results for subhalos abundances as a function of distance to host halo center and subhalo mass. Our results are in agreement with expectations for IDM models, namely we find a significantly smaller number of subhalos in IDM with respect to that observed in our CDM simulations. But not only the normalization of the cumulative subhalo mass function decreases (up to a factor $\sim$10 at the smallest resolved subhalo scales); also its slope is substantially lower in IDM ($\gamma=-0.71$ versus $\gamma=-0.88$ for CDM in the approximated range $10^{7}~M_{\odot}/h < m_{200} <  10^{9}~M_{\odot}/h$; see Fig.~\ref{fig:CSMF} and Tab.~\ref{tab:2}). As expected from theory, these differences among both cosmologies are not observed in the larger Box simulation. The radial distribution of subhalos within host halos exhibit a similar trend: there are fewer subhalos in IDM compared to CDM. Yet, we do not find appreciable differences in behaviour, i.e., the functional form of both radial distributions are similar.

In addition to the obvious interest for structure formation and study of halo and subhalo properties, we note that our work has a direct application on studies aimed at the indirect detection of DM, namely, the detection of the annihilation or decay products of DM particles. For instance, the extragalactic $\gamma$-ray and neutrino emission due DM annihilations depends mainly of the DM halos and subhalo properties~(see e.g.,~\cite{Ullio:2002,Ackermann:2015tah,Moline:2015,Moline:2017}).
Another example is the so-called {\it subhalo boost}: subhalos are expected to boost the  DM signal of their host halos significantly, e.g.~\cite{Sanchez-Conde:2014,Moline:2017}. This subhalo boost is very sensitive to the details of both subhalo concentration and subhalo abundance.
Overall, from our results we conclude that the role of halo substructure in DM searches will be less important in IDM scenarios than in CDM, given the fact that both the subhalo concentrations and abundances are lower in the former compared to the latter. Yet, it will not be not negligible, as we also find in our IDM simulations larger concentrations for subhalos with respect to field halos of the same mass. Although this work represents an important step on addressing this and related issues, a quantitative study of the precise role of IDM subhalos for DM searches is left for future work: the IDM cosmological model mainly impacts low mass structures, thus it will be necessary to have higher resolution simulations than those used in this work in order to do so. Likewise, for a full analysis of IDM halo and subhalo properties it will be also necessary to run IDM simulations adopting other values of the cross section of DM interactions.


\vspace{6pt}



\acknowledgments{AM, AAS, and MASC are supported by the {\it Atracci\'on de~Talento} Contract No. 2016-T1/TIC-1542 granted by the Comunidad de Madrid in Spain. They also acknowledge the support of the Spanish Agencia Estatal de Investigación through the grants PGC2018-095161-B-I00, IFT Centro de Excelencia Severo Ochoa SEV-2016-0597, and~Red Consolider MultiDark FPA2017-90566-REDC. AM also thanks the Institute of Astrophysics and Space Sciences of Portugal, where part of this work was done and the partial support of the RAICES Argentinian 
 program. We made use of the DiRAC
 Data Centric system at Durham University, operated by the ICC 
 on behalf of the STFC
 DiRAC HPC Facility (\texttt{www.dirac.ac.uk}). This equipment was funded by BIS 
 National E-infrastructure Capital Grant ST/K00042X/1, STFC
 Capital Grant ST/H008519/1, STFC DiRAC Operations Grant ST/K003267/1, and Durham University. DiRAC is part of the National E-Infrastructure. Furthermore, numerical computations were also done on the Sciama High Performance Compute (HPC) cluster, which is supported by the ICG
, SEPNet
, and the University of Portsmouth. SAC acknowledges funding from Consejo Nacional de Investigaciones
Cient\'{\i}ficas y T\'ecnicas (CONICET, PIP-0387), Agencia Nacional
de Promoci\'on Cient\'ifica y Tecnol\'ogica (ANPCyT, PICT-2013-0317), and Universidad Nacional de La Plata (G11-124; G11-150), Argentina.}


\conflictsofinterest{The authors declare no conflict of interest. The funders had no role in the design of the study; in the collection, analyses, or interpretation of data; in the writing of the manuscript, or in the decision to publish the results.}

\reftitle{References}

\end{document}